\documentclass[aps,twocolumn,floatfix]{revtex4-1}
\usepackage{physics}
\usepackage{bm} 
\usepackage{dcolumn} 
\usepackage{graphicx} 
\usepackage{amsmath} 
\usepackage{amssymb} 
\usepackage{amsfonts} 

\begin{document}
\title{Vertex corrections to the dc conductivity in anisotropic multiband systems}
\author{Sunghoon Kim}
\author{Seungchan Woo}
\author{Hongki Min}
\email{hmin@snu.ac.kr}
\affiliation{Department of Physics and Astronomy, Seoul National University, Seoul 08826, Korea}
\date{\today}

\begin{abstract}
For an isotropic single-band system, it is well known that the semiclassical Boltzmann transport theory within the relaxation time approximation and the Kubo formula with the vertex corrections provide the same result with the $(1-\cos\theta)$ factor in the inverse transport relaxation time. In anisotropic multiband systems, the semiclassical Boltzmann transport equation is generalized to coupled integral equations relating transport relaxation times at different angles in different bands. Using the Kubo formula, we study the vertex corrections to the dc conductivity in anisotropic multiband systems and derive the relation satisfied by the transport relaxation time for both elastic and inelastic scatterings, verifying that the result is consistent with the semiclassical approach.
\end{abstract}

\maketitle

\section{Introduction}

The semiclassical Boltzmann transport theory within the relaxation time approximation enables us to investigate the transport properties of various materials theoretically. For an isotropic system in which only a single band is involved in scattering, it is well known that the transport relaxation time $\tau_{\bm k}^{\rm tr}$ for a wavevector $\bm{k}$ in the relaxation time approximation can be expressed as \cite{Ashcroft1976}
\begin{equation}
\label{eq:relaxation_time_isotropic}
{1\over \tau_{\bm k}^{\rm tr}}=\int {d^d k' \over (2\pi)^d} W_{\bm{k}'\bm{k}} (1-\cos\theta_{\bm{k}\bm{k}'}),
\end{equation}
where $W_{{\bm k}'{\bm k}}$ is the transition rate from $\bm{k}$ state to $\bm{k}'$ state. The inverse relaxation time is a weighted average of the collision probability in which the forward scattering ($\theta_{\bm{k}\bm{k}'}=0$) receives reduced weight. 

\begin{figure}[htb]
\includegraphics[width=1\linewidth,height=1.7in]{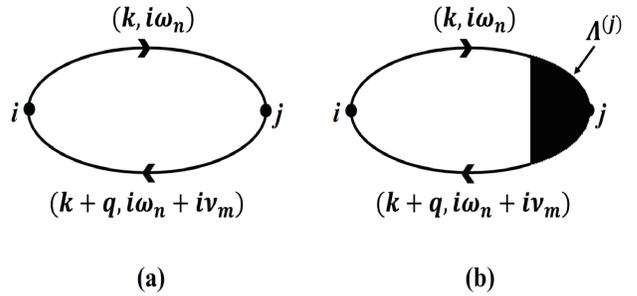}
\caption{
Schematic diagrams for (a) Drude conductivity without the $(1-\cos\theta)$ factor, (b) ladder diagrams giving the  $(1-\cos\theta)$ factor
} 
\label{fig:diagram}
\end{figure}

Using a many-body diagrammatic approach, the same result can be obtained from the current-current correlation functions supplemented with the ladder vertex corrections \cite{Mahan2000,Coleman2016}, as shown in Fig.~\ref{fig:diagram}. The single bubble diagram [Fig.~\ref{fig:diagram}(a)] captures the Drude conductivity with the quasiparticle lifetime $\tau_{\bm k}^{\rm qp}$, which does not contain the $(1-\cos\theta)$ factor, whereas the ladder diagrams [Fig.~\ref{fig:diagram}(b)] represent the leading-order corrections to the current vertex from impurity scattering, which give the $(1-\cos\theta)$ factor replacing $\tau_{\bm k}^{\rm qp}$ by the transport relaxation time $\tau_{\bm k}^{\rm tr}$ in Eq.~(\ref{eq:relaxation_time_isotropic}). While there are other sets of diagrams (for example, maximally crossed diagrams), the ladder diagrams are known to be dominant in the limit of small impurity density.

In anisotropic multiband systems, the relaxation time in the semiclassical Boltzmann theory is not simply given by Eq.~(\ref{eq:relaxation_time_isotropic}); rather its relation is generalized to coupled integral equations relating the relaxation times at different angles in different bands \cite{Xiao2016, Brosco2016, Xiao2017, Siggia1970, Vyborny2009, Breitkreiz2013, Liu2016, Park2017,Woo2017,Park2018}. Many materials, such as nodal line semimetals \cite{Burkov2011,Fang2016}, multi-Weyl semimetals \cite{Fang2012}, and few-layer black phosphorus \cite{Fei2015,Carvalho2016,Rui2017}, have a Fermi surface that is anisotropic or crosses multiple energy bands. Thus, it is important to include the effects of the anisotropy and multiple energy bands of the system to describe its transport properties correctly.

There have been numerous theoretical studies on the equivalence between the semiclassical Boltzmann transport equation and the Kubo formula
\cite{Langer1960, Prange1964, Holstein1964, Hansch1983, Cappelluti2009}. As in the case of isotropic single-band systems, the semiclassical Boltzmann approach and the many-body diagrammatic approach are expected to provide consistent results for anisotropic multiband systems \cite{Prange1964, Holstein1964}. However, most of the previous works have focused on isotropic single-band systems without considering anisotropic Fermi surfaces or multiple energy bands at the Fermi energy, and there has been no systematic study demonstrating the relation between the two approaches in anisotropic multiband systems, which will provide a firm foundation for the transport theory in these systems. Although the equivalence between two approaches has been conjectured by many researchers, the rigorous proof without assuming any single-band or isotropic nature has been challenging.

In this work, we study the vertex corrections to the dc conductivity in anisotropic multiband systems in the weak scattering limit. Using a diagrammatic method and the corresponding Kubo formula, we derive the equivalent results for both elastic and inelastic scatterings obtained from the semiclassical approach. For elastic scattering, we consider randomly distributed impurities in the limit of small impurity density. For inelastic scattering, we consider the electron-phonon interaction. For both cases, we do not assume a specific form of scattering potential. We also prove that the Ward identities are satisfied in anisotropic multiband systems.

The rest of this paper is organized as follows. In Sec. \ref{sec:semiclassical_approach}, we briefly review the semiclassical Boltzmann transport theory in anisotropic multiband systems for both elastic and inelastic scatterings. In Sec. \ref{sec:diagramatic_approach}, using a diagrammatic approach, we develop a theory for the vertex corrections in anisotropic multiband systems, proving that the diagrammatic approach gives the same result as the semiclassical approach. We conclude in Sec.~\ref{sec:discussion and summary} with a discussion on the Ward identities. In Appendix \ref{app:momentum_integral_first}, we present an alternative diagrammatic approach by performing the momentum integral first rather than the frequency summation first used in the main text. In Appendixes \ref{app:freqency_sum_first} and \ref{app:Ward_identities}, we provide detailed derivations for the vertex corrections and the Ward identities, respectively, which were omitted in the main text.

\section{Semiclassical approach}
\label{sec:semiclassical_approach}
In this section, we provide the semiclassical approach to calculate transport properties. We use the semiclassical Boltzmann theory within the first-order Born approximation, which is known to be valid in the weak scattering limit \cite{Kohn1957, Ando1982, Ferry2009}.

\subsection{Elastic scattering}
In this section, we briefly review the semiclassical Boltzmann theory in \(d\)-dimensional anisotropic multiband systems for elastic scattering. In the following derivation, we assume that electrons are scattered from randomly distributed impurities. In this study, we set the reduced Planck constant \(\hbar\) to 1 for convenience.

Let \(f(\bm{r},\bm{k};t)\) denote the distribution function of an electron at the state \(\bm{k}\) at position \(\bm{r}\) at time \(t\). The rate of change of \(f(\bm{r},\bm{k};t)\) with respect to time satisfies the following equation:
\begin{equation}
\label{eq:collision integral}
    \frac{df}{dt}=\frac{\partial f}{\partial \bm{r}}\cdot \bm{v}_{\bm{k}} +\frac{\partial f}{\partial \bm{k}}\cdot \bm{\dot \bm{k}}+\frac{\partial f}{\partial t}, 
\end{equation}
where \(\bm{v}_{\bm{k}}\) is the velocity at the state \(\bm{k}\). Assuming a homogeneous system without the explicit time-dependence in \(f(\bm{r},\bm{k};t)\equiv f_{\bm k}\), Eq.~(\ref{eq:collision integral}) reduces to  $\frac{df}{dt}=\frac{\partial f_{\bm{k}}}{\partial \bm{k}}\cdot \bm{\dot \bm{k}}$. 
In the presence of collision, the collision integral is given by

\begin{eqnarray}
\label{eq:collision integral2}
\left(\frac{df}{dt}\right)_{\rm c}\!\!\!=\!\int\! \frac{d^dk'}{(2\pi)^d}\left[W_{\bm{k}\bm{k}'}f_{\bm{k}'}(1-f_{\bm{k}})\!-\!W_{\bm{k}'\bm{k}}f_{\bm{k}}(1-f_{\bm{k}'})\right],\nonumber\\
\end{eqnarray}
where \(W_{\bm{k}'\bm{k}}\) is the transition rate from \(\bm{k}\) to \(\bm{k}'\). The first term in the right-hand side of Eq.~(\ref{eq:collision integral2}) describes the probability per unit time that an electron is scattered into a state $\bm k$ and the second term describes the probability per unit time that an electron in a state $\bm k$ is scattered out.
The Boltzmann transport equation is given by 
\(\left(\frac{df}{dt}\right)\)=\(\left(\frac{df}{dt}\right)_{\rm c}\).
Expanding the theory to multiband systems, the Boltzmann transport equation can be generalized as
\begin{eqnarray}
\label{eq:collision integral3}
  \frac{\partial f_{\alpha,\bm{k}}}{\partial \bm{k}}\cdot \bm{\dot \bm{k}}&=&\sum_{\alpha'}\int \frac{d^dk'}{(2\pi)^d}\{W_{\alpha,\bm{k};\alpha',\bm{k}'}f_{\alpha',\bm{k}'}(1-f_{\alpha,\bm{k}})\nonumber\\
  &&\,\,\, -W_{\alpha',\bm{k}';\alpha,\bm{k}}f_{\alpha,\bm{k}}(1-f_{\alpha',\bm{k}'})\},
\end{eqnarray}
where \(\alpha\) and \(\alpha'\) are band indices. 

For elastic scattering, the transition rate \(W_{\alpha',\bm{k}';\alpha,\bm{k}}\) is given by the following form:
\begin{equation}
\label{eq:transition_rate_elastic}
W_{\alpha',\bm{k}';\alpha,\bm{k}}=2\pi n_{\rm imp} |V_{\alpha',\bm{k}';\alpha,\bm{k}}|^2 \delta(\xi_{\alpha,\bm{k}}- \xi_{\alpha',\bm{k}'}),
\end{equation}
where \(n_{\rm imp}\) is the impurity density, \(V_{\alpha',\bm{k}';\alpha,\bm{k}}=\matrixel{\alpha',\bm{k}'}{V}{\alpha,\bm{k}}\) is the matrix element of the impurity potential \(V\), which describes a scattering from \((\alpha,\bm{k})\) to \((\alpha',\bm{k}')\), and \(\xi_{\alpha,\bm{k}}\equiv \varepsilon_{\alpha,\bm{k}}-\mu\) is the energy of an electron at the state \((\alpha,\bm{k})\) measured from the chemical potential \(\mu\). Here, the effect of electron-electron interactions can be taken into account through the screening of the impurity potential. Note that \(W_{\alpha,\bm{k};\alpha',\bm{k}'}=W_{\alpha',\bm{k}';\alpha,\bm{k}}\); thus, Eq.~(\ref{eq:collision integral3}) reduces to
\begin{eqnarray}
\label{eq:collision integral4}
  \frac{\partial f_{\alpha,\bm{k}}}{\partial \bm{k}}\cdot \bm{\dot \bm{k}}=\sum_{\alpha'}\int \frac{d^dk'}{(2\pi)^d} W_{\alpha',\bm{k}';\alpha,\bm{k}}(f_{\alpha',\bm{k}'}-f_{\alpha,\bm{k}}).
\end{eqnarray}
In the presence of a small external electric field, we assume that \(f_{\alpha,\bm{k}}\) deviates slightly from \(f_{\alpha,\bm{k}}^{0}\):
\begin{equation}
   f_{\alpha,\bm{k}}= f_{\alpha,\bm{k}}^{0}+\delta f_{\alpha,\bm{k}},  
\end{equation}
where \(f_{\alpha,\bm{k}}^{0}\equiv f^{0}(\xi_{\alpha,\bm{k}})=\left[e^{\beta \xi_{\alpha,\bm{k}}}+1\right]^{-1}\) is the Fermi--Dirac distribution function in equilibrium with $\beta={1\over k_{\rm B}T}$. We assume that the deviation \(\delta f_{\alpha,\bm{k}}\) can be parameterized up to first order of the electric field $\bm{E}$ as follows \cite{Park2017,Woo2017,Park2018}:
\begin{equation}
\label{eq:deviation of f}
  \delta f_{\alpha,\bm{k}}=(-e)\sum_{i}E^{(i)}v_{\alpha,\bm{k}}^{(i)}\tau_{\alpha,\bm{k}}^{(i)}S^{0}(\xi_{\alpha,\bm{k}}),
\end{equation}
where \(S^{0}(\xi)=-\frac{\partial f^{0}(\xi)}{\partial \xi}=\beta f^{0}(\xi)\left[1-f^{0}(\xi)\right]\), and \(v_{\alpha,\bm{k}}^{(i)}\) and \(\tau_{\alpha,\bm{k}}^{(i)}\) are the velocity and transport relaxation time along the \(i\)th direction at the state \((\alpha,\bm{k})\), respectively. Inserting Eq.~(\ref{eq:deviation of f}) into Eq.~(\ref{eq:collision integral4}), we have an integral equation relating the relaxation times at different states
\begin{equation}
\label{eq:relaxation_time_anisotropic_multiband}
1=\sum_{\alpha'}\int {d^d k'\over (2\pi)^d} W_{\alpha',\bm{k}';\alpha,\bm{k}}\left(\tau_{\alpha,\bm{k}}^{(i)}-{v_{\alpha',\bm{k}'}^{(i)}\over v_{\alpha,\bm{k}}^{(i)} }\tau_{\alpha',\bm{k}'}^{(i)}\right).
\end{equation}
Note that, for an isotropic single-band system, Eq.~(\ref{eq:relaxation_time_anisotropic_multiband}) reduces to Eq.~(\ref{eq:relaxation_time_isotropic}). 

The deviation of the electron distribution function from the equilibrium value gives rise to the current density
\begin{equation}
\label{eq:current tensor}
  J^{(i)}=g\sum_{\alpha}\int \frac{d^dk}{(2\pi)^d}(-e)v_{\alpha,\bm{k}}^{(i)}\delta f_{\alpha,\bm{k}}=\sum_{j}\sigma_{ij}E^{(j)},
\end{equation}
where \(g\) is the degeneracy factor and \(\sigma_{ij}\) is a matrix element of the conductivity tensor given by
\begin{equation}
\label{eq:conductivity tensor}
  \sigma_{ij}=ge^2 \sum_{\alpha}\int  \frac{d^dk}{(2\pi)^d}S^{0}(\xi_{\alpha,\bm{k}}) v_{\alpha,\bm{k}}^{(i)}v_{\alpha,\bm{k}}^{(j)} \tau_{\alpha,\bm{k}}^{(j)}.
\end{equation}

\subsection{Inelastic scattering}
For inelastic scattering, such as phonon-mediated scattering, Eq.~(\ref{eq:relaxation_time_anisotropic_multiband}) is no longer valid and the principle of detailed balance should be considered  \cite{Kawamura1992}:
\begin{equation}
\label{eq:the principle of detailed balance}
 W_{\alpha',\bm{k}'; \alpha,\bm{k}}f_{\alpha,\bm{k}}^{0}(1-f_{\alpha',\bm{k}'}^{0})=W_{\alpha,\bm{k}; \alpha',\bm{k}'}f_{\alpha',\bm{k}'}^{0}(1-f_{\alpha,\bm{k}}^{0}). 
\end{equation}
Expanding up to first order of \(\delta f\), Eq.~(\ref{eq:collision integral3}) reduces to
\begin{eqnarray}
\label{eq:collision integral_inelastic}
  \frac{\partial f_{\alpha,\bm{k}}^{0}}{\partial \bm{k}}\cdot \bm{\dot \bm{k}}&=&\sum_{\alpha'}\int \frac{d^dk'}{(2\pi)^d}W_{\alpha',\bm{k}';\alpha,\bm{k}}
  \nonumber\\&&\times \left(\frac{f_{\alpha,\bm{k}}^{0}}{f_{\alpha',\bm{k}'}^{0}}\delta f_{\alpha',\bm{k}'}-\frac{1-f_{\alpha',\bm{k}'}^{0}}{f_{\alpha,\bm{k}}^{0}}\delta f_{\alpha,\bm{k}}\right).
\end{eqnarray}
Using the parameterization in Eq.~(\ref{eq:deviation of f}), we obtain an integral equation for inelastic scattering relating the relaxation times at different states as follows:
\begin{eqnarray}
\label{eq:relaxation time_inelastic}
  1&=&\sum_{\alpha'}\int {d^d k'\over (2\pi)^d} W_{\alpha',\bm{k}';\alpha,\bm{k}}
  \nonumber\\&&\times \left(\tau_{\alpha,\bm{k}}^{(i)}-{v_{\alpha',\bm{k}'}^{(i)}\over v_{\alpha,\bm{k}}^{(i)} }\tau_{\alpha',\bm{k}'}^{(i)}\right) \left(\frac{1-f_{\alpha',\bm{k}'}^{0}}{1-f_{\alpha,\bm{k}}^{0}}\right).
\end{eqnarray}
Note that the integral equation for inelastic scattering is different from that for elastic scattering by the factor \(\frac{1-f_{\alpha',\bm{k}'}^{0}}{1-f_{\alpha,\bm{k}}^{0}}\). 

For phonon scattering, the transition rate \(W_{\alpha',\bm{k}';\alpha,\bm{k}}\) is given by \cite{Ziman1960}
\begin{eqnarray}
\label{eq:phonon transition rate}
    W_{\alpha',\bm{k}';\alpha,\bm{k}}&=&2\pi \sum_{\lambda}|\matrixel{\alpha', \bm{k}'}{M_\lambda}{\alpha,\bm{k}}|^2 \nonumber\\
    &&\times\{\left[n_{\rm B}(\Omega_{\lambda,\bm{q}})+1\right]\delta(\xi_{\alpha',\bm{k}'}-\xi_{\alpha,\bm{k}}+\Omega_{\lambda,\bm{q}}) \nonumber\\
    &&+n_{\rm B}(\Omega_{\lambda,\bm{q}})\delta(\xi_{\alpha',\bm{k}'}-\xi_{\alpha,\bm{k}}-\Omega_{\lambda,\bm{q}})\},
\end{eqnarray}
where \(n_{\rm B}(\Omega_{\lambda,\bm{q}})=\left[e^{\beta\Omega_{\lambda,\bm{q}}}-1\right]^{-1}\) is the Bose--Einstein distribution function, and \(M_{\lambda}\) denotes the electron-phonon interaction for the phonon polarization \(\lambda\). Here, the first (second) term on the right-hand side of Eq.~(\ref{eq:phonon transition rate}) describes the emission (absorption) of a phonon with momentum \(\bm{q}=\pm(\bm{k}-\bm{k}')\) and frequency \(\Omega_{\lambda,\bm{q}}\). In this work, umklapp processes are neglected since we are interested in the weak scattering limit where normal processes are dominant \cite{Bruus2004}.

\section{Diagrammatic approach}
\label{sec:diagramatic_approach}
In this section, using a diagrammatic approach, we develop a theory for the vertex corrections to the dc conductivity for elastic and inelastic scatterings in \(d\)-dimensional anisotropic multiband systems, and verify that the results are consistent with those obtained from the semiclassical Boltzmann equation in Sec.~\ref{sec:semiclassical_approach}. 

The dc conductivity is obtained by taking the long wavelength limit and then the static limit as follows \cite{Mahan2000}:
\begin{eqnarray}
\label{eq:dc conductivity delta function}
    \sigma_{ij}^{\rm dc}&&=-\lim_{\nu\rightarrow 0}\frac{1}{\nu} {\rm Im} \Pi_{ij}(\bm{q}=0,\nu),
\end{eqnarray}
where \(\Pi_{ij}(\bm{q},\nu)\) is the retarded current-current response function, which is obtained using the analytic continuation \(i\nu_m\rightarrow \nu+i0^{+}\) of the current-current response function \(\Pi_{ij}(\bm{q},i\nu_m)\) for an imaginary frequency. First, we consider the single bubble diagram without the vertex corrections [Fig.~\ref{fig:diagram}(a)]
\begin{eqnarray}
\label{eq:current-current response function}
    \Pi_{ij}(i\nu_m)\!&=&\!\frac{ge^2}{\beta \mathcal{V}}\!\!\!\sum_{\alpha,\alpha',\bm{k},i\omega_n}\!\! \mathcal{G}_{\alpha}(\bm{k},i\omega_n)v^{(i)}_{\alpha,\alpha'}(\bm{k},\bm{k}) 
    \nonumber\\&&\times \mathcal{G}_{\alpha'}(\bm{k},i\omega_n+i\nu_m)v^{(j)}_{\alpha',\alpha}(\bm{k}),
\end{eqnarray}
where \(\Pi_{ij}(i\nu_m)=\Pi_{ij}(\bm{q}=0,i\nu_m)\), \(\omega_n\) and \(\nu_m\) are fermionic and bosonic Matsubara frequencies, respectively,
\(\mathcal{V}\) is the volume of the system, \(\mathcal{G}_{\alpha}(\bm{k},i\omega_n)\) is the interacting Green's function, and \(v^{(j)}_{\alpha',\alpha}(\bm{k})=\matrixel{\alpha',\bm{k}}{\hat{v}^{(j)}}{\alpha,\bm{k}}\) is the matrix element of the velocity operator \(\hat{v}^{(j)}=\frac{\partial \hat H}{\partial k_j}\) along the \(j\)th direction. The velocity matrix element can be expressed as
\begin{eqnarray}
\label{eq:velocity operator}
v^{(j)}_{\alpha',\alpha}(\bm{k})&=&v^{(j)}_{\alpha,\bm{k}}\braket{\alpha',\bm{k}}{\alpha,\bm{k}}
\nonumber\\&&+(\varepsilon_{\alpha,\bm{k}}-\varepsilon_{\alpha',\bm{k}})\matrixel{\alpha',\bm{k}}{\frac{\partial}{\partial k_j}}{\alpha,\bm{k}}.
\end{eqnarray} 
In the \(\nu\rightarrow 0\) limit, the second term in Eq.~(\ref{eq:velocity operator}) does not contribute to \(\Pi_{ij}(i\nu_m)\) as finite energy transfer between \((\alpha',\bm{k})\) and \((\alpha,\bm{k})\) is not allowed in the single bubble diagram. By choosing an orthonormal basis set, the right-hand side of Eq.~(\ref{eq:velocity operator}) simply reduces to \(v^{(j)}_{\alpha,\bm{k}}\delta_{\alpha',\alpha}\), and only diagonal elements of the velocity matrix remain in Eq.~(\ref{eq:current-current response function}). 

Incorporating the ladder diagrams, we finally obtain the current-current response function at low frequencies supplemented with the vertex corrections as follows [Fig.~\ref{fig:diagram}(b)]:
\begin{eqnarray}
\label{eq:current-current response3}
  \Pi_{ij}(i\nu_m)&&=\frac{ge^2}{\beta \mathcal{V}}\sum_{\alpha,\bm{k},i\omega_n}\mathcal{G}_{\alpha}(\bm{k},i\omega_n)v^{(i)}_{\alpha,\bm{k}}\mathcal{G}_{\alpha}(\bm{k},i\omega_n+i\nu_m)
  \nonumber\\&&\times v^{(j)}_{\alpha,\bm{k}}\Lambda^{(j)}_{\alpha}(\bm{k},i\omega_n,i\omega_n+i\nu_m),
\end{eqnarray}
where \(v^{(j)}_{\alpha,\bm{k}}\Lambda^{(j)}_{\alpha}(\bm{k},i\omega_n,i\omega_n+i\nu_m)\) is the vertex corresponding to the current density operator along the \(j\)th direction. Note that we only included the diagonal elements of the velocity matrix, as discussed above.

To calculate the dc conductivity using a many-body diagrammatic method, we can either perform the Matsubara frequency summation first or the momentum integral first. Here, we use the former method where the frequency summation is performed first, and present the other method in Appendix \ref{app:momentum_integral_first}.

\subsection{Elastic scattering}
For elastic scattering, we consider randomly distributed impurities. The effect of impurities can be considered using the disorder-averaged Green's function
\begin{equation}
\label{eq:disorder-averaged Green's function}
   \mathcal{G}_{\alpha}(\bm{k},i\omega_{n})=\frac{1}{i\omega_{n}-\xi_{\alpha,\bm{k}}-\Sigma_{\alpha}(\bm{k},i\omega_n)},
\end{equation}
where \(\Sigma_{\alpha}(\bm{k},i\omega_n)\) is the electron self-energy from impurity scattering. The imaginary part of the self-energy can be associated with the quasiparticle lifetime $\tau_{\alpha,\bm{k}}^{\rm qp}$ as ${\rm Im}\Sigma_{\alpha}(\bm{k},i\omega_n)=-{1\over 2 \tau_{\alpha,\bm{k}}^{\rm qp}}{\rm sgn}(\omega_n)$. Assuming small impurity density, the inverse of the quasiparticle lifetime is given by
\begin{eqnarray}
\label{eq:quasiparticle_lifetime}
\frac{1}{\tau_{\alpha,\bm{k}}^{\rm qp}}
&=&2\pi n_{\rm imp} \sum_{\alpha'} \int \frac{d^dk'}{(2\pi)^d}  |V_{\alpha',\bm{k}';\alpha,\bm{k}}|^2 \delta(\xi_{\alpha,\bm{k}}- \xi_{\alpha',\bm{k}'}) \nonumber \\
&=&\sum_{\alpha'} \int \frac{d^dk'}{(2\pi)^d} W_{\alpha',\bm{k}';\alpha,\bm{k}}.
\end{eqnarray}
Note that the integrand in the right-hand side of the first line is identical to \(W_{\alpha',\bm{k}';\alpha,\bm{k}}\) defined in Eq.~(\ref{eq:transition_rate_elastic}).

\begin{figure}[htb]
\includegraphics[width=1\linewidth,height=2in]{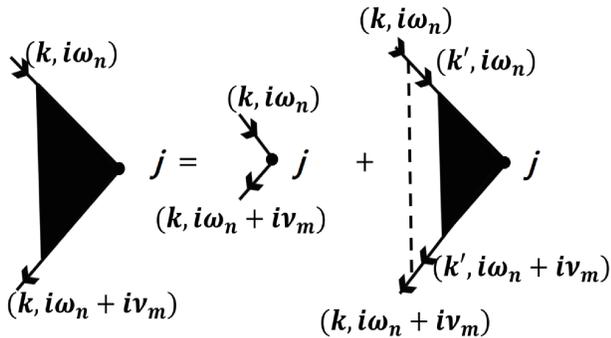}
\caption{
Diagrams for the ladder vertex corrections for elastic scattering
} 
\label{fig:vertex}
\end{figure}

Within the ladder approximation, the vertex correction is approximated by a sum of ladder diagrams given by the self-consistent Dyson form as follows [Fig.~\ref{fig:vertex}]:
\begin{eqnarray}
\label{eq:vertex_dyson's equation}
&&v_{\alpha,\bm{k}}^{(j)}\Lambda^{(j)}_{\alpha}(\bm{k},i\omega_n,i\omega_n+i\nu_m)  \\
&=&v_{\alpha,\bm{k}}^{(j)} +\frac{n_{\rm imp}}{\mathcal{V}} \sum_{\alpha',\bm{k}'}|V_{\alpha',\bm{k}';\alpha,\bm{k}}|^2 \mathcal{G}_{\alpha'}(\bm{k}',i\omega_n)
\nonumber \\ 
 &&\times v_{\alpha',\bm{k}'}^{(j)}\Lambda^{(j)} _{\alpha'}(\bm{k}',i\omega_n,i\omega_n+i\nu_m) \mathcal{G}_{\alpha'}(\bm{k}',i\omega_n +i\nu_m). \nonumber
\end{eqnarray}
As the form of the self-consistent Dyson's equation in Eq.~(\ref{eq:vertex_dyson's equation}) is analogous to Eq.~(\ref{eq:relaxation_time_anisotropic_multiband}), \(\Lambda^{(j)}\) can be related to the transport relaxation time. Here, we derive this relation rigorously.

\begin{figure}[htb]
\includegraphics[width=0.8\linewidth,height=2.3in]{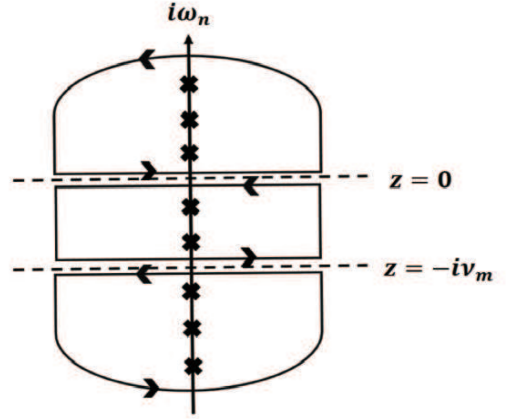}
\caption{
Contour \(C\) used for the contour integral in Eq.~(\ref{eq:current-current response function2}) having two branch cuts along the axes \(z=0\) and \(z=-i\nu_m\)
} 
\label{fig:contour integral}
\end{figure}

Let us first calculate the current-current response function. Eq.~(\ref{eq:current-current response3}) can be expressed as the frequency sum of the following form:
\begin{eqnarray}
\label{eq:current-current response function2}
\Pi_{ij}(i\nu_m)&\equiv&\frac{1}{\beta}\sum_{i\omega_n}P(i\omega_n,i\omega_n+i\nu_m) \nonumber \\
&=&-\oint_{C} \frac{dz}{2\pi i} \frac{P(z,z+i\nu_m)}{e^{\beta z}+1},
\end{eqnarray}
where \(P(i\omega_n,i\omega_n+i\nu_m)\) is a complex function whose summation can be performed via integration along the contour $C$ shown in Fig.~\ref{fig:contour integral}. Note that the contour integral in Eq.~(\ref{eq:current-current response function2}) has poles at $z=i\omega_n$.

After performing the analytic continuation $i\nu_m\rightarrow \nu+i0^{+}$ and assuming small impurity density, we finally obtain the dc conductivity as (see Appendix \ref{app:freqency_sum_first})
\begin{eqnarray}
\label{eq:conductivity_integral}
    \sigma_{ij}&=&\int \frac{d\xi}{2\pi} S^{0}(\xi) P^{\rm{AR}}(\xi,\xi),
\end{eqnarray}
where \(P^{\rm{AR}}(\xi,\xi)=P(\xi-i0^{+},\xi+i0^{+})\). Henceforth, the superscripts $\rm A$ and $\rm R$ represent advanced and retarded functions, respectively. Note that
\begin{eqnarray}
\label{eq:P_AR}
&&P^{\rm{AR}}(\xi,\xi) \\
&=&\frac{g e^2}{\mathcal{V}}\sum_{\alpha,\bm{k}} v_{\alpha,\bm{k}}^{(i)} G_{\alpha}^{\rm A}(\bm{k},\xi) v_{\alpha,\bm{k}}^{(j)} \Lambda^{(j){\rm{AR}}}_{\alpha}(\bm{k},\xi,\xi) G_{\alpha}^{\rm R}(\bm{k},\xi) \nonumber \\
&=&\frac{2\pi g e^2}{\mathcal{V}}\sum_{\alpha,\bm{k}} v_{\alpha,\bm{k}}^{(i)} v_{\alpha,\bm{k}}^{(j)} \tau_{\alpha,\bm{k}}^{\rm qp} \delta(\xi-\xi_{\alpha,\bm{k}}) \Lambda^{(j){\rm{AR}}}_{\alpha}(\bm{k},\xi,\xi), \nonumber 
\end{eqnarray}
where $G_{\alpha}^{\rm R,A}(\bm{k},\xi)= \mathcal{G}_{\alpha}(\bm{k},i\omega_n\rightarrow \xi\pm i0^{+}$).
Here, we used $G_{\alpha}^{\rm A}(\bm{k},\xi)G_{\alpha}^{\rm R}(\bm{k},\xi)\rightarrow 2\pi\tau^{\rm qp}_{\alpha,\bm{k}}\delta(\xi-\xi_{\alpha,\bm{k}})$ in the \(n_{\rm imp}\rightarrow 0\) limit.

Therefore, from Eqs.~(\ref{eq:conductivity_integral}) and (\ref{eq:P_AR}), the dc conductivity is given by
\begin{eqnarray}
\label{eq:dc conductivity}
\sigma_{ij}&=&\frac{ge^2}{\mathcal{V}} \sum_{\alpha,\bm{k}} S^0(\xi_{\alpha,\bm{k}}) v_{\alpha,\bm{k}}^{(i)} v_{\alpha,\bm{k}}^{(j)} \tau_{\alpha,\bm{k}}^{\rm qp} \Lambda^{(j){\rm{AR}}}_{\alpha}(\bm{k},\xi_{\alpha,\bm{k}},\xi_{\alpha,\bm{k}}). \nonumber \\
\end{eqnarray}
For detailed derivations, see Appendix \ref{app:freqency_sum_first}.

Comparing Eq.~(\ref{eq:dc conductivity}) with Eq.~(\ref{eq:conductivity tensor}), it is natural to relate \(\Lambda^{(j)}\) to the transport relaxation time along the corresponding direction. To determine this relation, let us return to the self-consistent Dyson's equation for the vertex correction in Eq.~(\ref{eq:vertex_dyson's equation}). After analytic continuation, it reduces to
\begin{eqnarray}
\label{eq:vertex_dyson's equation2}
&&\Lambda^{(j){\rm{AR}}}_{\alpha}(\bm{k},\xi,\xi) \\
&=&1+\frac{n_{\rm imp}}{\mathcal{V}} \sum_{\alpha',\bm{k}'}|V_{\alpha',\bm{k}';\alpha,\bm{k}}|^2\frac{v_{\alpha',\bm{k}'}^{(j)}}{v_{\alpha,\bm{k}}^{(j)}} \nonumber\\
&&\,\,\,\,\,\times G^{\rm A}_{\alpha'}(\bm{k}',\xi) \Lambda^{(j){\rm{AR}}}_{\alpha'}(\bm{k}',\xi,\xi) G^{\rm R}_{\alpha'}(\bm{k}',\xi) \nonumber\\ 
&=&1+\sum_{\alpha'} \int \frac{d^dk'}{(2\pi)^d} W_{\alpha',\bm{k}';\alpha,\bm{k}} \frac{v_{\alpha',\bm{k}'}^{(j)}}{v_{\alpha,\bm{k}}^{(j)}} \tau_{\alpha',\bm{k}'}^{\rm qp}\Lambda^{(j){\rm{AR}}}_{\alpha'}(\bm{k}',\xi,\xi). \nonumber
\end{eqnarray}
Let us define the transport relaxation time along the \(j\)th direction as
\begin{equation}
\label{eq:transport relaxation time}
\tau_{\alpha,\bm{k}}^{(j)} \equiv \tau_{\alpha,\bm{k}}^{\rm qp} \Lambda^{(j){\rm{AR}}} _{\alpha}(\bm{k},\xi_{\alpha,\bm{k}},\xi_{\alpha,\bm{k}}).
\end{equation}
Subsequently, Eq.~(\ref{eq:vertex_dyson's equation2}) can be rewritten as
\begin{eqnarray}
\label{eq:vertex_dyson's equation3}
\frac{\tau_{\alpha,\bm{k}}^{(j)}}{\tau_{\alpha,\bm{k}}^{\rm qp}}&=&1+\sum_{\alpha'} \int \frac{d^dk'}{(2\pi)^d} W_{\alpha',\bm{k}';\alpha,\bm{k}} \frac{v_{\alpha',\bm{k}'}^{(j)}}{v_{\alpha,\bm{k}}^{(j)}} \tau_{\alpha',\bm{k}'}^{(j)}.
\end{eqnarray}
Using the definition of the quasiparticle lifetime in Eq.~(\ref{eq:quasiparticle_lifetime}), we obtain an integral equation for the transport relaxation time for elastic scattering in anisotropic multiband systems as
\begin{eqnarray}
\label{eq:relaxation time_anisotropic_elastic2}
1&=& \sum_{\alpha'} \int \frac{d^dk'}{(2\pi)^d} W_{\alpha',\bm{k}';\alpha,\bm{k}} \left(\tau_{\alpha,\bm{k}}^{(j)}-\frac{v_{\alpha',\bm{k}'}^{(j)}}{v_{\alpha,\bm{k}}^{(j)}}\tau_{\alpha',\bm{k}'}^{(j)}\right),
\end{eqnarray}
which is the same as the semiclassical result in Eq.~(\ref{eq:relaxation_time_anisotropic_multiband}). Furthermore, using the definition of the transport relaxation time, we can easily verify that Eq.~(\ref{eq:dc conductivity}) is consistent with Eq.~(\ref{eq:conductivity tensor}) obtained from the semiclassical approach.

\subsection{Inelastic scattering}
As in the case of elastic scattering, we develop a theory for the vertex corrections for inelastic scattering. Here, we specifically consider phonon-mediated scattering, which yields intrinsic resistivity in a metal.

\begin{figure}[htb]
\includegraphics[width=1\linewidth,height=1.7in]{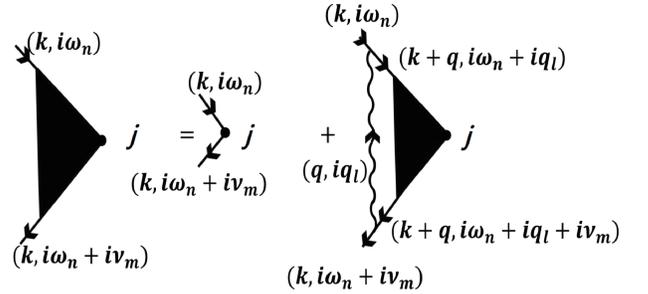}
\caption{
Dyson's equation for the ladder vertex corrections for inelastic scattering
} 
\label{fig:vertex phonon}
\end{figure}

The self-consistent Dyson's equation of the vertex part for phonon scattering is given by [Fig.~\ref{fig:vertex phonon}] 
\begin{eqnarray}
\label{eq:dyson's equation_phonon}
    &&v_{\alpha,\bm{k}}^{(j)}\Lambda^{(j)}_{\alpha}(\bm{k},i\omega_n,i\omega_n +i\nu_m)
    \nonumber\\
    &&=v_{\alpha,\bm{k}}^{(j)}-\frac{1}{\beta \mathcal{V}}\sum_{\alpha',\bm{q},iq_l,\lambda}|\matrixel{\alpha',\bm{k}+\bm{q}}{M_\lambda}{\alpha,\bm{k}}|^2 D_{\lambda}(\bm{q},iq_l)
    \nonumber\\
    &&\times  \mathcal{G}_{\alpha'}(\bm{k}+\bm{q},i\omega_n +iq_l)\mathcal{G}_{\alpha'}(\bm{k}+\bm{q},i\omega_n+iq_l + i\nu_m  ) 
    \nonumber\\
    &&\times v_{\alpha',\bm{k}+\bm{q}}^{(j)}\Lambda^{(j)}_{\alpha'}(\bm{k}+\bm{q},i\omega_n +iq_l,i\omega_n+iq_l+i\nu_m ), 
\end{eqnarray}
where \(q_l\) is a bosonic Matsubara frequency and \(D_{\lambda}(\bm{q},iq_l)=\frac{2\Omega_{\lambda,\bm{q}}}{(iq_l)^2-\Omega_{\lambda,\bm{q}}^2}\) is the non-interacting phonon Green's function with the renormalized phonon frequency \(\Omega_{\lambda,\bm{q}}\). Eq.~(\ref{eq:dyson's equation_phonon}) can be rewritten as
\begin{eqnarray}
\label{eq:dyson's equation_phonon2}
   &&\Lambda^{(j)}_{\alpha}(\bm{k},i\omega_n,i\omega_n +i\nu_m) \\
   &&=1-\sum_{\alpha',\lambda}\int \frac{d^dq}{(2\pi)^d}|\matrixel{\alpha',\bm{k}+\bm{q}}{M_\lambda}{\alpha,\bm{k}}|^2 \frac{v_{\alpha',\bm{k}+\bm{q}}^{(j)}}{v_{\alpha,\bm{k}}^{(j)}} \nonumber\\
   &&\times \frac{1}{\beta} \sum_{iq_l} Q(iq_l+i\omega_n,iq_l+i\omega_n+i\nu_m), \nonumber 
\end{eqnarray}
where
\begin{eqnarray}
&&Q(iq_l+i\omega_n,iq_l+i\omega_n+i\nu_m)\\
&\equiv&\mathcal{G}_{\alpha'}(\bm{k}+\bm{q},iq_l+i\omega_n)\mathcal{G}_{\alpha'}(\bm{k}+\bm{q},iq_l+i\omega_n+ i\nu_m) \nonumber\\
&& \times D_{\lambda}(\bm{q},iq_l)\Lambda^{(j)}_{\alpha'}(\bm{k}+\bm{q},iq_l+i\omega_n,iq_l+i\omega_n+i\nu_m). \nonumber
\end{eqnarray}

\begin{figure}[htb]
\includegraphics[width=0.9\linewidth,height=2.3in]{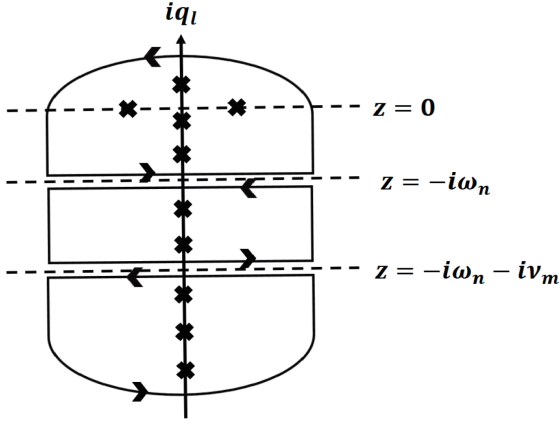}
\caption{
Contour \(C'\) used for the contour integral in Eq.~(\ref{eq:contour_integral_C'}) having two branch cuts along the axes \(z=-i\omega_n\) and \(z=-i\omega_{n}-i\nu_{m}\)
} 
\label{fig:contour integral_phonon}
\end{figure}

The summation of \(Q(iq_l+i\omega_n,iq_l+i\omega_n+i\nu_m)\) over the bosonic Matsubara frequency \(q_l\)  can be performed with the aid of a contour integral along the contour \(C'\) shown in Fig.~\ref{fig:contour integral_phonon}:
\begin{eqnarray}
\label{eq:contour_integral_C'}
\oint_{C'} \frac{dz}{2\pi i} \frac{Q(z+i\omega_n,z+i\omega_n+i\nu_m)}{e^{\beta z}-1}.
\end{eqnarray}
Note that the contour integral in Eq.~(\ref{eq:contour_integral_C'}) has poles at $z=iq_l$ as well as $z=\pm \Omega_{\lambda,\bm{q}}$.

According to Eq.~(\ref{eq:dc conductivity}), the only vertex part contributing to the dc conductivity is \(\Lambda^{(j)\rm{AR}}(\bm{k},\xi_{\alpha,\bm{k}},\xi_{\alpha,\bm{k}})\). After performing the analytic continuation \(i\omega_n \rightarrow \xi-i0^{+}\) and \(i\omega_n+i\nu_m \rightarrow \xi +\nu +i0^{+}\) in Eq.~(\ref{eq:dyson's equation_phonon2}), 
and assuming weak scattering, the self-consistent Dyson's equation at $\nu=0$ reduces to 
\begin{eqnarray}
\label{eq:dyson_phonon}
 &&\Lambda^{(j)\rm{AR}}_{\alpha}(\bm{k},\xi,\xi) 
 \\
 &&=1+2\pi\sum_{\alpha',\lambda}\int \frac{d^dq}{(2\pi)^d}|\matrixel{\alpha',\bm{k}+\bm{q}}{M_\lambda}{\alpha,\bm{k}}|^2\frac{v_{\alpha',\bm{k}+\bm{q}}^{(j)}}{v_{\alpha,\bm{k}}^{(j)}} \nonumber\\
 && \times \tau_{\alpha',\bm{k}+\bm{q}}^{\rm qp} \Lambda^{(j)\rm{AR}}_{\alpha'}(\bm{k}+\bm{q},\xi_{\alpha',\bm{k}+\bm{q}},\xi_{\alpha',\bm{k}+\bm{q}}) \nonumber\\
 &&\times\{\left[n_{\rm B}(\Omega_{\lambda,\bm{q}})+f^0(\xi+\Omega_{\lambda,\bm{q}})\right]\delta(\xi + \Omega_{\lambda,\bm{q}} -\xi_{\alpha',\bm{k}+\bm{q}}) \nonumber\\
 &&+\left[n_{\rm B}(\Omega_{\lambda,\bm{q}})+1-f^0(\xi-\Omega_{\lambda,\bm{q}})\right]\delta(\xi - \Omega_{\lambda,\bm{q}} -\xi_{\alpha',\bm{k}+\bm{q}})\}. \nonumber
\end{eqnarray}
For detailed derivations, see Appendix \ref{app:freqency_sum_first}.

Finally, let us replace \(\Lambda^{(j)}\) by the transport relaxation time defined in Eq.~(\ref{eq:transport relaxation time}). The quasiparticle lifetime in the presence of phonon scattering is given by \cite{Ziman1960}
\begin{eqnarray}
\label{eq:qusiparticle life time_phonon}
    \frac{1}{\tau_{\alpha,\bm{k}}^{\rm qp}}&=&2\pi \sum_{\alpha',\lambda} \int \frac{d^dq}{(2\pi)^d}|\matrixel{\alpha',\bm{k}+\bm{q}}{M_\lambda}{\alpha,\bm{k}}|^2 \\
    &\times&     \{\left[1+n_{\rm B}(\Omega_{\lambda,\bm{q}})-f^{0}_{\alpha',\bm{k}+\bm{q}}\right]\delta(\xi_{\alpha,\bm{k}}-\xi_{\alpha,\bm{k}+\bm{q}}-\Omega_{\lambda,\bm{q}})\nonumber\\
    &+&\left[n_{\rm B}(\Omega_{\lambda,\bm{q}})+f^{0}_{\alpha',\bm{k}+\bm{q}}\right]\delta(\xi_{\alpha,\bm{k}}-\xi_{\alpha',\bm{k}+\bm{q}}+\Omega_{\lambda,\bm{q}})\}. \nonumber
\end{eqnarray}
Here, we replaced \(f^0(\xi_{\alpha',\bm{k}+\bm{q}})\) by \(f^{0}_{\alpha',\bm{k}+\bm{q}}\).

Notably, for phonon emission process \((\xi_{\alpha,\bm{k}}=\xi_{\alpha',\bm{k}+\bm{q}}+\Omega_{\lambda,\bm{q}})\),
\begin{eqnarray}
\label{idensity_phonon_emission}
1+n_{\rm B}(\Omega_{\lambda,\bm{q}})-f^{0}_{\alpha',\bm{k}+\bm{q}}=\left[1+n_{\rm B}(\Omega_{\lambda,\bm{q}})\right]\left(\frac{1-f^{0}_{\alpha',\bm{k}+\bm{q}}}{1-f^{0}_{\alpha,\bm{k}}}\right), \nonumber\\
\end{eqnarray}
whereas for phonon absorption process \((\xi_{\alpha,\bm{k}}=\xi_{\alpha',\bm{k}+\bm{q}}-\Omega_{\lambda,\bm{q}})\),
\begin{eqnarray}
\label{idensity_phonon_absorption}
n_{\rm B}(\Omega_{\lambda,\bm{q}})+f^{0}_{\alpha',\bm{k}+\bm{q}}=n_{\rm B}(\Omega_{\lambda,\bm{q}})\left(\frac{1-f^{0}_{\alpha',\bm{k}+\bm{q}}}{1-f^{0}_{\alpha,\bm{k}}}\right).
\end{eqnarray}
Thus, Eq.~(\ref{eq:qusiparticle life time_phonon}) can be rewritten as
\begin{eqnarray}
\label{eq:qusiparticle life time_phonon2}
\frac{1}{\tau_{\alpha,\bm{k}}^{\rm qp}}=
\sum_{\alpha'}\int \frac{d^dq}{(2\pi)^d} W_{\alpha',\bm{k}+\bm{q};\alpha,\bm{k}}\left(\frac{1-f^{0}_{\alpha',\bm{k}+\bm{q}}}{1-f^{0}_{\alpha,\bm{k}}}\right),
\end{eqnarray}
where \(W_{\alpha',\bm{k}+\bm{q};\alpha,\bm{k}}\) is the transition rate defined in Eq.~(\ref{eq:phonon transition rate}).

Using Eqs.~(\ref{idensity_phonon_emission}) and (\ref{idensity_phonon_absorption}), and replacing \(\Lambda^{(j)}\) by the transport relaxation time, we can rewrite Eq.~(\ref{eq:dyson_phonon}) as
\begin{eqnarray}
\label{eq:dyson_phonon2}
 \frac{\tau_{\alpha,\bm{k}}^{(j)}}{\tau_{\alpha,\bm{k}}^{\rm qp}}=1&+&\sum_{\alpha'}\int \frac{d^dq}{(2\pi)^d} W_{\alpha',\bm{k}+\bm{q};\alpha,\bm{k}}
 \nonumber\\
 &&\times \frac{v_{\alpha',\bm{k}+\bm{q}}^{(j)}}{v_{\alpha,\bm{k}}^{(j)}} \tau_{\alpha',\bm{k}+\bm{q}}^{(j)}
 \left(\frac{1-f^{0}_{\alpha',\bm{k}+\bm{q}}}{1-f^{0}_{\alpha,\bm{k}}}\right).
\end{eqnarray}
Using the definition of quasiparticle lifetime in Eq.~(\ref{eq:qusiparticle life time_phonon2}), we finally obtain an integral equation for the transport relaxation time for inelastic scattering in anisotropic multiband systems given by
\begin{eqnarray}
  1&=&\sum_{\alpha'}\int {d^d k'\over (2\pi)^d} W_{\alpha',\bm{k}+\bm{q};\alpha,\bm{k}}
  \nonumber\\&&\times \left(\tau_{\alpha,\bm{k}}^{(j)}-{v_{\alpha',\bm{k}+\bm{q}}^{(j)}\over v_{\alpha,\bm{k}}^{(j)} }\tau_{\alpha',\bm{k}+\bm{q}}^{(j)}\right) \left(\frac{1-f_{\alpha',\bm{k}+\bm{q}}^{0}}{1-f_{\alpha,\bm{k}}^{0}}\right),
\end{eqnarray}
which is consistent with the semiclassical result in Eq.~(\ref{eq:relaxation time_inelastic}).

\section{Discussion and summary}
\label{sec:discussion and summary}

The validity of the diagrammatic approach for the vertex corrections in this work can be verified by testing the Ward identity \cite{Ward1950, Engelsberg1963}. The Ward identity is the exact relationship between the self-energy and the vertex correction arising from the continuity equation, which must hold if the corresponding diagrams are properly included both in the self-energy and the vertex correction.
For both elastic and inelastic scatterings, we demonstrate that the Ward identity is satisfied in anisotropic multiband systems as follows (see Appendix \ref{app:Ward_identities}):
\begin{eqnarray}
v_{\alpha,\bm{k}}^{(j)}\Lambda^{(j)}_{\alpha}(\bm{k},i\omega_n,i\omega_n)=v^{(j)}_{\alpha,\bm{k}}+\frac{\partial\Sigma_{\alpha}(\bm{k},i\omega_n)}{\partial k^{(j)}},
\end{eqnarray}
which indicates that we have employed the proper vertex part \(\Lambda_{\alpha}^{(j)}(\bm{k},i\omega_n,i\omega_n+i\nu_m)\) corresponding to the ladder self-energy diagrams.

The ladder approximation is known to be valid in the limit of small impurity density or weak scattering. As the impurity density increases, further terms called the maximally crossed diagrams, which arise from the coherent interference between electron wave functions, should be incorporated into the vertex corrections, leading to the quantum corrections to the conductivity called the weak localization.

In summary, using a many-body diagrammatic approach, we studied the vertex corrections to the dc conductivity in anisotropic multiband systems, demonstrating that the diagrammatic approach provides an equivalent result to that obtained from the semiclassical Boltzmann approach for both elastic and inelastic scatterings. Our work provides a many-body justification for the generalized Boltzmann transport theory given by coupled integral equations for anisotropic multiband systems, which is essential to capture the effects of the anisotropy and multiple energy bands in transport correctly.

\acknowledgments
We thank Ki-Seok Kim for useful discussions. This work was supported by the NRF grant funded by the Korea government (MSIT) (No. 2018R1A2B6007837) and Creative-Pioneering Researchers Program through Seoul National University (SNU).

\appendix

\section{Alternative derivations for the vertex corrections}
\label{app:momentum_integral_first}
In this section, following chapter 10 of Coleman \cite{Coleman2016}, we derive the vertex corrections to the dc conductivity for elastic impurity scattering by performing the momentum integral first. Let us first consider the self-consistent Dyson's equation in Eq.~(\ref{eq:vertex_dyson's equation}). The electrons on the Fermi surface mainly contribute to the dc conductivity. Therefore, we focus on calculating the vertex corrections for electrons at the Fermi energy. As the two Green's functions on the right-hand side of Eq.~(\ref{eq:vertex_dyson's equation}) become appreciable near the Fermi energy at low frequencies, we separate the two terms from the rest. Thus, Eq.~(\ref{eq:vertex_dyson's equation}) reduces to 
\begin{eqnarray}
\label{eq:dyson_elastic_kfirst}
&&\Lambda^{(j)}_{\alpha}(\bm{k},i\omega_n,i\omega_n+i\nu_m) \nonumber\\ 
&& \approx 1+n_{\rm imp} \sum_{\alpha'}\int\frac{d^dk'}{(2\pi)^d}|V_{\alpha',\bm{k}';\alpha,\bm{k}}|^2 \frac{v_{\alpha',\bm{k}'}^{(j)}}{v_{\alpha,\bm{k}}^{(j)}}
 \nonumber\\&&\times \Lambda^{(j)}_{\alpha'}(\bm{k}',i\omega_n,i\omega_n+i\nu_m)\delta(\xi_{\alpha',\bm{k}'})
 \nonumber\\ &&\times \int d\xi \mathcal{G}(\xi,i\omega_n)\mathcal{G}(\xi,i\omega_n +i\nu_m),
\end{eqnarray}
where
\begin{equation}
\label{eq:coleman_green's function}
  \mathcal{G}(\xi,i\omega_n)=\frac{1}{i\omega_{n}-\xi+i{\rm sgn}(\omega_n)\frac{1}{2\tau^{\rm{qp}}_{\alpha,\bm{k}}}}
\end{equation}
is the disorder-averaged Green's function up to the first-order Born approximation [see Eq.~(\ref{eq:disorder-averaged Green's function}) in the main text]. Note that impurity scattering of electrons at the Fermi energy provides a constant contribution to the real part of the self-energy, which can be absorbed in the chemical potential.

Let us calculate the energy integral first. We can compute the energy integral in Eq.~(\ref{eq:dyson_elastic_kfirst}) with the aid of a contour integral method as follows:
\begin{eqnarray}
\label{eq:energy integral}
\int d\xi \mathcal{G}(\xi,i\omega_n)\mathcal{G}(\xi,i\omega_n +i\nu_m)
= \frac{2\pi \Theta(\nu_m,\omega_n)}{\nu_m+\frac{1}{\tau_{\alpha',\bm{k}'}^{\rm{qp}}}},
\end{eqnarray}
where
\begin{equation}
\Theta(\nu_m,\omega_n)=
\left\{ \begin{array}{cl} 
1& \mbox{for $-\nu_m<\omega_n<0$,} \\ 
0 & \mbox{otherwise}.
\end{array}\right.
\end{equation}
Here, we assumed $\nu_m>0$. Thus, Eq.~(\ref{eq:dyson_elastic_kfirst}) reduces to
\begin{eqnarray}
\label{eq:dyson_elastic_kfirst2}
&&\Lambda^{(j)}_{\alpha}(\bm{k},i\omega_n,i\omega_n+i\nu_m) \nonumber\\ 
&& \approx 1+n_{\rm imp} \sum_{\alpha'}\int\frac{d^dk'}{(2\pi)^d} \delta(\xi_{\alpha',\bm{k}'}) |V_{\alpha',\bm{k}';\alpha,\bm{k}}|^2 \frac{v_{\alpha',\bm{k}'}^{(j)}}{v_{\alpha,\bm{k}}^{(j)}}
 \nonumber\\
 &&\,\,\, \times \Lambda^{(j)}_{\alpha'}(\bm{k}',i\omega_n,i\omega_n+i\nu_m) \frac{2\pi\Theta(\nu_m,\omega_n)}{\nu_m+\frac{1}{\tau_{\alpha',\bm{k}'}^{\rm{qp}}}}.
\end{eqnarray}
Here, the integral provides a non-zero value only if the poles of the two Green's functions are on the opposite sides with respect to the real axis in frequency space.

Note that, because of the \(\Theta(\nu_m,\omega_n)\) term, \(\Lambda^{(j)}_{\alpha}(\bm{k},i\omega_n,i\omega_n+i\nu_m)\) has a value independent of \(\omega_n\) within the range  \(-\nu_{m}<\omega_{n}<0\), and otherwise $1$. Thus, \(\Lambda^{(j)}_{\alpha}(\bm{k},i\omega_n,i\omega_n+i\nu_m)\) can be expressed as
\begin{eqnarray}
&&\Lambda^{(j)}_{\alpha}(\bm{k},i\omega_n,i\omega_n+i\nu_m) \nonumber\\
&&=\left\{ \begin{array}{cl} 
\Lambda^{(j)} _{\alpha}(\bm{k},i\nu_m)& \mbox{ for $-\nu_m<\omega_n<0$,} \\ 
1 & \mbox{ otherwise.}
\end{array}\right.
\end{eqnarray}
Here, we assumed \(\Lambda^{(j)}_{\alpha}(\bm{k},i\omega_n,i\omega_n+i\nu_m)=\Lambda^{(j)} _{\alpha}(\bm{k},i\nu_m)\) for \(-\nu_m<\omega_n<0\).

An alternative expression of the dc conductivity in the imaginary time formalism is given by 
\begin{eqnarray}
\label{eq:dc conductivity_Coleman}
  \sigma_{ij}(i\nu_m)=\frac{1}{\nu_m}\left[\Pi_{ij}(i\nu_m)-\Pi_{ij}(0)\right].
\end{eqnarray}
Note that electrons near the Fermi surface mostly contribute to the difference between the current-current response functions at $\nu_m$ and $\nu_m=0$ \cite{Coleman2016}. Therefore, the dc conductivity can be obtained as
\begin{widetext}
\begin{eqnarray}
\label{eq:dc conductivity_coleman}
    \sigma_{ij}(i\nu_m)&&=\frac{ge^2}{\beta \nu_m \mathcal{V}}\sum_{\alpha,\bm{k},i\omega_n}\left[v_{\alpha,\bm{k}}^{(i)}v_{\alpha,\bm{k}}^{(j)} \Lambda^{(j)}_{\alpha}(\bm{k},i\omega_n,i\omega_n+i\nu_m)\mathcal{G}_{\alpha}(\bm{k},i\omega_n)\mathcal{G}_{\alpha}(\bm{k},i\omega_n+i\nu_m) -(i\nu_{m}\rightarrow 0)\right] \nonumber\\
    &&\approx \frac{ge^2}{\beta \nu_m}\sum_{\alpha,i\omega_n} \int \frac{d^dk}{(2\pi)^d}\delta(\xi_{\alpha,\bm{k}}) v_{\alpha,\bm{k}}^{(i)}v_{\alpha,\bm{k}}^{(j)}\left[\Lambda^{(j)}_{\alpha}(i\omega_n,i\omega_n+i\nu_m)\int d\xi\mathcal{G}(\xi,i\omega_n)\mathcal{G}(\xi,i\omega_n+i\nu_m)-(i\nu_m \rightarrow 0)\right]  \nonumber\\
    &&=\frac{ge^2}{\beta \nu_m}\sum_\alpha \int \frac{d^dk}{(2\pi)^d}\delta(\xi_{\alpha,\bm{k}}) v_{\alpha,\bm{k}}^{(i)}v_{\alpha,\bm{k}}^{(j)} \sum_{i\omega_n}\Lambda^{(j)}_{\alpha}(\bm{k},i\omega_n,i\omega_n+i\nu_m)\frac{2\pi}{\nu_m+\frac{1}{\tau_{\alpha,\bm{k}}^{\rm{qp}}}}\Theta(\nu_m,\omega_n) \nonumber \\
    &&=ge^2\sum_\alpha \int \frac{d^dk}{(2\pi)^d}\delta(\xi_{\alpha,\bm{k}}) v_{\alpha,\bm{k}}^{(i)}v_{\alpha,\bm{k}}^{(j)} \frac{\Lambda^{(j)}_{\alpha}(\bm{k},i\nu_m)}{\nu_m+\frac{1}{\tau_{\alpha,\bm{k}}^{\rm{qp}}}}.    
\end{eqnarray}
\end{widetext}
Here, we used \(\frac{1}{\beta}\sum_{\omega_n} \Theta(\nu_m,\omega_n)=\frac{\nu_m}{2\pi}\). Therefore, by defining the transport relaxation time along the \(j\)th direction as
\begin{eqnarray}
\label{eq:transport relaxation time_coleman}
    \tau_{\alpha,\bm{k}}^{(j)}
\equiv \lim_{\nu_m\rightarrow0}\Lambda^{(j)}_{\alpha}(\bm{k},i\nu_m)\tau_{\alpha,\bm{k}}^{\rm{qp}},
\end{eqnarray}
we obtain a result consistent with the dc conductivity obtained through the semiclassical approach in Eq.~(\ref{eq:conductivity tensor}). 

Finally, let us perform \(\lim_{\nu_m\rightarrow0}\frac{1}{\beta}\sum_{\omega_n}\Theta(\nu_m,\omega_n)\) on both sides of Eq.~(\ref{eq:dyson_elastic_kfirst2}), followed by multiplication by \(\frac{2\pi}{\nu_m}\). 
Thus, we have
\begin{eqnarray}
\frac{\tau_{\alpha,\bm{k}}^{(j)}}{\tau_{\alpha,\bm{k}}^{\rm{qp}}}
&=&1+2\pi n_{\rm imp} \sum_{\alpha'}\int\frac{d^dk'}{(2\pi)^d}|V_{\alpha',\bm{k}';\alpha,\bm{k}}|^2 \nonumber\\
&&\,\,\, \times \delta(\xi_{\alpha,\bm{k}}- \xi_{\alpha',\bm{k}'}) \frac{v_{\alpha',\bm{k}'}^{(j)}}{v_{\alpha,\bm{k}}^{(j)}} \tau_{\alpha',\bm{k}'}^{(j)}.
\end{eqnarray}
Here, we assumed \( \xi_{\alpha,\bm{k}}=\xi_{\alpha',\bm{k}'}\approx0\). 
Therefore, we have an integral equation relating the transport relaxation times as follows:
\begin{eqnarray}
  1&=&2\pi n_{\rm imp}\sum_{\alpha'}\int \frac{d^dk'}{(2\pi)^d} |V_{\alpha',\bm{k}';\alpha,\bm{k}}|^2
  \nonumber\\&&\times \delta(\xi_{\alpha,\bm{k}}- \xi_{\alpha',\bm{k}'}) \left(\tau_{\alpha,\bm{k}}^{(j)}-\frac{v_{\alpha',\bm{k}'}^{(j)}}{v_{\alpha,\bm{k}}^{(j)}}\tau_{\alpha',\bm{k}'}^{(j)}\right),
\end{eqnarray}
which is consistent with the semiclassical result in Eq.~(\ref{eq:relaxation_time_anisotropic_multiband}).

\section{Detailed derivations for the vertex corrections}
\label{app:freqency_sum_first}
\subsection{Elastic scattering}
In this section, following chapter 8 of Mahan \cite{Mahan2000}, we present detailed derivations for the vertex corrections for elastic scattering. 
Let us first consider the contour integral in Eq.~(\ref{eq:current-current response function2}):
\begin{widetext}
\begin{eqnarray}
\Pi_{ij}(i\nu_m)&=&-\oint_{C} \frac{dz}{2\pi i} f^{0}(z)P(z,z+i\nu_m)  \\
&=&\int \frac{d\xi}{2\pi i}f^{0}(\xi) \left[-P(\xi+i0^{+},\xi+i\nu_m)+P(\xi-i0^{+},\xi+i\nu_m)
-P(\xi-i\nu_m,\xi+i0^{+})+P(\xi-i\nu_m,\xi-i0^{+})\right].\nonumber
\end{eqnarray}
\end{widetext}
After performing the analytic continuation \((i\nu_{m}\rightarrow \nu+i0^{+})\), we have
\begin{eqnarray}
\Pi_{ij}^{\rm R}(\nu)&&=
\int \frac{d\xi}{2\pi i} \{[f^{0}(\xi)-f^{0}(\xi+\nu)]P^{\rm AR}(\xi,\xi+\nu) \\
&&\,\,\, -f^{0}(\xi)P^{\rm RR}(\xi,\xi+\nu)+f^{0}(\xi+\nu)P^{\rm AA}(\xi,\xi+\nu)\}. \nonumber
\end{eqnarray}
Thus, in the \(\nu\rightarrow 0\) limit, the dc conductivity can be rewritten as
\begin{eqnarray}
    \sigma_{ij}&=&\frac{ge^2}{2\pi}\int d\xi S^{0}(\xi)
    [P^{\rm{AR}}(\xi,\xi)-{\rm Re}P^{\rm{RR}}(\xi,\xi)],
\end{eqnarray}
which includes the \(P^{\rm AR}(\xi,\xi)\) and \(P^{\rm RR}(\xi,\xi)\) terms in the integrand. 

Here, we show that only the first term on the right-hand side contributes to the dc conductivity whereas the second term becomes negligible in the limit of small impurity density. 
Before computing each term, we note several useful formulas pertaining to the spectral function \(A_{\alpha}(\bm{k},\xi)=-2{\rm Im} G^{\rm{R}}_{\alpha}(\bm{k},\xi)\): 
\begin{subequations}
\label{eq:spectral_approx}
\begin{eqnarray}
\lim_{\Delta_{\alpha,\bm{k}}\rightarrow0}A_{\alpha}(\bm{k},\xi)&=&2\pi\delta(\xi-\xi_{\alpha,\bm{k}}), \\
\lim_{\Delta_{\alpha,\bm{k}}\rightarrow0}A^2_{\alpha}(\bm{k},\xi)&=&\frac{2\pi\delta(\xi-\xi_{\alpha,\bm{k}})}{\Delta_{\alpha,\bm{k}}},
\end{eqnarray}
\end{subequations}
where \(\Delta_{\alpha,\bm{k}}\equiv \frac{1}{2\tau^{\rm qp}_{\alpha,\bm{k}}}\). Note that, in the \(\Delta_{\alpha,\bm{k}}\rightarrow 0\) limit, or equivalently in the \(n_{\rm imp} \rightarrow 0\) limit, the spectral function reduces to a delta function. 

First, let us evaluate the contribution of the \(P^{\rm{RR}}(\xi,\xi)\) term to the dc conductivity:
\begin{eqnarray}
    &&P^{\rm{RR}}(\xi,\xi) \\
    &&=\frac{ge^2}{\mathcal{V}}\sum_{\alpha,\bm{k}}v_{\alpha,\bm{k}}^{(i)}G_{\alpha}^{\rm{R}}(\bm{k},\xi) v_{\alpha,\bm{k}}^{(j)} 
    \Lambda^{(j)\rm{RR}}_{\alpha}(\bm{k},\xi,\xi)G_{\alpha}^{\rm{R}}(\bm{k},\xi). \nonumber
\end{eqnarray}
In the \(n_{\rm imp} \rightarrow 0\) limit, the product of the two Green's functions vanishes and the contribution of \(P^{\rm RR}(\xi,\xi)\) to the dc conductivity becomes negligible \cite{Mahan2000}. 

Subsequently, let us compute the \(P^{\rm{AR}}(\xi,\xi)\) term as follows:
\begin{eqnarray}
    &&P^{\rm{AR}}(\xi,\xi) \\
    &&=\frac{ge^2}{\mathcal{V}}\sum_{\alpha,\bm{k}}v_{\alpha,\bm{k}}^{(i)}G^{\rm{A}}_{\alpha}(\bm{k},\xi) v_{\alpha,\bm{k}}^{(j)}\Lambda^{(j)\rm{AR}}_{\alpha}(\bm{k},\xi,\xi) G^{\rm{R}}_{\alpha}(\bm{k},\xi) \nonumber \\
    &&=\frac{2\pi ge^2}{\mathcal{V}}\sum_{\alpha,\bm{k}} v_{\alpha,\bm{k}}^{(i)} v_{\alpha,\bm{k}}^{(j)} \tau^{\rm qp}_{\alpha,\bm{k}}\delta(\xi-\xi_{\alpha,\bm{k}})
    \Lambda^{(j)\rm{AR}}_{\alpha}(\bm{k},\xi,\xi).
    \nonumber
\end{eqnarray}
Therefore, the dc conductivity can be simplified as
\begin{eqnarray}
    \sigma_{ij}&&=\frac{1}{2\pi}\int d\xi S^0(\xi)P^{\rm AR}(\xi,\xi)
    \\&&=\frac{ge^2}{\mathcal{V}}\sum_{\alpha,\bm{k}}S^{0}(\xi_{\alpha,\bm{k}})v_{\alpha,\bm{k}}^{(i)} v_{\alpha,\bm{k}}^{(j)}\tau^{\rm qp}_{\alpha,\bm{k}}\Lambda^{(j)\rm{AR}}_{\alpha}(\bm{k},\xi_{\alpha,\bm{k}},\xi_{\alpha,\bm{k}}),\nonumber
\end{eqnarray}
thus yielding Eq.~(\ref{eq:dc conductivity}) in the main text.

\begin{widetext}
\subsection{Inelastic scattering}
In this section, we present detailed derivations for the vertex corrections for inelastic scattering. Let us first consider Eq.~(\ref{eq:contour_integral_C'}) in the main text. The contour integral along \(C'\) has two types of poles: \(z=iq_{l}\) and \(z=\pm \Omega_{\lambda,\bm{q}}\). Therefore, the summation \(\mathcal{S}(i\omega_n,i\omega_n+i\nu_m)\equiv \frac{1}{\beta}\sum_{iq_l}Q(iq_l+i\omega_n,iq_l+i\omega_n+i\nu_m)\) can be rewritten as follows:
\begin{eqnarray}
\label{eq:matsubara sum Q}
&&\mathcal{S}(i\omega_n,i\omega_n+i\nu_m) \nonumber\\
&&=\oint_{C'} \frac{dz}{2\pi i} n_{\rm B}(z) Q(z+i\omega_n,z+i\omega_n+i\nu_m) \nonumber \\
&&-n_{\rm{B}}(\Omega_{\lambda,\bm{q}})\mathcal{G}_{\alpha'}(\bm{k}+\bm{q},i\omega_n+\Omega_{\lambda,\bm{q}}+ i\nu_m) \mathcal{G}_{\alpha'}(\bm{k}+\bm{q},i\omega_n +\Omega_{\lambda,\bm{q}})\Lambda^{(j)}_{\alpha'}(\bm{k}+\bm{q},i\omega_n+\Omega_{\lambda,\bm{q}},i\omega_n+\Omega_{\lambda,\bm{q}}+i\nu_m)  \\
&&-[n_{\rm{B}}(\Omega_{\lambda,\bm{q}})+1]\mathcal{G}_{\alpha'}(\bm{k}+\bm{q},i\omega_n-\Omega_{\lambda,\bm{q}}+ i\nu_m)
   \mathcal{G}_{\alpha'}(\bm{k}+\bm{q},i\omega_n -\Omega_{\lambda,\bm{q}})\Lambda^{(j)}_{\alpha'}(\bm{k}+\bm{q},i\omega_n-\Omega_{\lambda,\bm{q}},i\omega_n-\Omega_{\lambda,\bm{q}}+i\nu_m),\nonumber
\end{eqnarray}
where the contour integral can be decomposed as
\begin{eqnarray}
\label{eq:contour integral C'_2}
   &&\oint_{C'}\frac{dz}{2\pi i}n_{\rm B}(z)Q(z+i\omega_n,z+i\omega_n+i\nu_m)
   \nonumber\\&&=-\int \frac{d\xi'}{2\pi i} f^{0}(\xi')\frac{2\Omega_{\lambda,\bm{q}}}{(\xi'-i\omega_n)^2-\Omega_{\lambda,\bm{q}}^2}
   \{\mathcal{G}_{\alpha'}(\bm{k}+\bm{q},\xi'+i0^{+})\mathcal{G}_{\alpha'}(\bm{k}+\bm{q},\xi'+i\nu_m)
   \Lambda^{(j)}_{\alpha'}(\bm{k}+\bm{q},\xi'+i0^{+},\xi'+i\nu_m)
   \nonumber\\&&-\mathcal{G}_{\alpha'}(\bm{k}+\bm{q},\xi'-i0^{+})\mathcal{G}_{\alpha'}(\bm{k}+\bm{q},\xi'+i\nu_m)
   \Lambda^{(j)}_{\alpha'}(\bm{k}+\bm{q},\xi'-i0^{+},\xi'+i\nu_m)\}
   \nonumber\\
   &&-\int \frac{d\xi'}{2\pi i} f^{0}(\xi')\frac{2\Omega_{\lambda,\bm{q}}}{(\xi'-i\omega_n-i\nu_m)^2-\Omega_{\lambda,\bm{q}}^2}
   \{\mathcal{G}_{\alpha'}(\bm{k}+\bm{q},\xi'-i\nu_m)\mathcal{G}_{\alpha'}(\bm{k}+\bm{q},\xi'+i0^{+})
   \Lambda^{(j)}_{\alpha'}(\bm{k}+\bm{q},\xi'-i\nu_m,\xi'+i0^{+})
   \nonumber\\
   &&-\mathcal{G}_{\alpha'}(\bm{k}+\bm{q},\xi'-i\nu_m)\mathcal{G}_{\alpha'}(\bm{k}+\bm{q},\xi'-i0^{+})
   \Lambda^{(j)}_{\alpha'}(\bm{k}+\bm{q},\xi'-i\nu_m,\xi'-i0^{+})\}.
\end{eqnarray}
To compute \(\Lambda^{(j)\rm{AR}}(\bm{k},\xi,\xi)\), let us perform the analytic continuation \(i\omega_n \rightarrow \xi-i0^{+}\) and \(i\omega_n+i\nu_m \rightarrow \xi +\nu +i0^{+}\). Thus, Eq.~(\ref{eq:matsubara sum Q}) at \(\nu=0\) reduces to
\begin{eqnarray}
\mathcal{S}^{\rm{AR}}(\xi,\xi)&=&-n_{\rm{B}}(\Omega_{\lambda,\bm{q}})|G^{\rm{R}}_{\alpha'}(\bm{k}+\bm{q},\xi+\Omega_{\lambda,\bm{q}})|^2 \Lambda^{(j)\rm{AR}}_{\alpha'}(\bm{k}+\bm{q},\xi+\Omega_{\lambda,\bm{q}},\xi+\Omega_{\lambda,\bm{q}})
\\&&
-[1+n_{\rm{B}}(\Omega_{\lambda,\bm{q}})]|G^R_{\alpha'}(\bm{k}+\bm{q},\xi-\Omega_{\lambda,\bm{q}})|^2
\Lambda^{(j)\rm{AR}}_{\alpha'}(\bm{k}+\bm{q},\xi-\Omega_{\lambda,\bm{q}},\xi-\Omega_{\lambda,\bm{q}})
\nonumber\\&&-\int \frac{d\xi'}{2\pi i} f^{0}(\xi')|G^{\rm{R}}_{\alpha'}(\bm{k}+\bm{q},\xi')|^2
\Lambda^{(j)\rm{AR}}_{\alpha'}(\bm{k}+\bm{q},\xi',\xi') \left[\frac{2\Omega_{\lambda,\bm{q}}}{(\xi'-\xi-i0^{+})^2-\Omega_{\lambda,\bm{q}}^2}-\frac{2\Omega_{\lambda,\bm{q}}}{(\xi'-\xi+i0^{+})^2-\Omega_{\lambda,\bm{q}}^2}\right].\nonumber
\end{eqnarray}
The last integration over \(\xi'\) can be performed with the aid of the Cauchy principal value
\begin{eqnarray}
\frac{2\Omega_{\lambda,\bm{q}}}{(\xi'-\xi-i0^{+})^2-\Omega_{\lambda,\bm{q}}^2}-\frac{2\Omega_{\lambda,\bm{q}}}{(\xi'-\xi+i0^{+})^2-\Omega_{\lambda,\bm{q}}^2}=2\pi i \left[\delta(\xi'-\xi-\Omega_{\lambda,\bm{q}})-\delta(\xi'-\xi+\Omega_{\lambda,\bm{q}})\right].
\end{eqnarray}
Therefore, in the weak-scattering limit, the self-consistent Dyson's equation for inelastic scattering [Eq.~(\ref{eq:dyson's equation_phonon})] can be rewritten as
\begin{eqnarray}
 &&\Lambda^{(j)\rm{AR}}_{\alpha}(\bm{k},\xi,\xi) \\
 &&=1-\sum_{\alpha',\lambda}\int \frac{d^dq}{(2\pi)^d}|\matrixel{\alpha',\bm{k}+\bm{q}}{M_\lambda}{\alpha,\bm{k}}|^2 \frac{v_{\alpha',\bm{k}+\bm{q}}^{(j)}}{v_{\alpha,\bm{k}}^{(j)}}\mathcal{S}^{\rm{AR}}(\xi,\xi) \nonumber\\
 &&=1+2\pi\sum_{\alpha',\lambda}\int \frac{d^dq}{(2\pi)^d}|\matrixel{\alpha',\bm{k}+\bm{q}}{M_\lambda}{\alpha,\bm{k}}|^2\frac{v_{\alpha',\bm{k}+\bm{q}}^{(j)}}{v_{\alpha,\bm{k}}^{(j)}} \tau_{\alpha',\bm{k}+\bm{q}}^{\rm qp} \Lambda^{(j)\rm{AR}}_{\alpha'}(\bm{k}+\bm{q},\xi_{\alpha',\bm{k}+\bm{q}},\xi_{\alpha',\bm{k}+\bm{q}}) \nonumber\\
 &&\,\,\, \times\{\left[n_{\rm B}(\Omega_{\lambda,\bm{q}})+f^0(\xi+\Omega_{\lambda,\bm{q}})\right]\delta(\xi + \Omega_{\lambda,\bm{q}} -\xi_{\alpha',\bm{k}+\bm{q}}) +\left[n_{\rm B}(\Omega_{\lambda,\bm{q}})+1-f^0(\xi-\Omega_{\lambda,\bm{q}})\right]\delta(\xi - \Omega_{\lambda,\bm{q}} -\xi_{\alpha',\bm{k}+\bm{q}})\},\nonumber 
\end{eqnarray}

thus yielding Eq.~(\ref{eq:dyson_phonon}) in the main text.
\end{widetext}

\begin{figure}[htb]
\includegraphics[width=1\linewidth,height=2.1in]{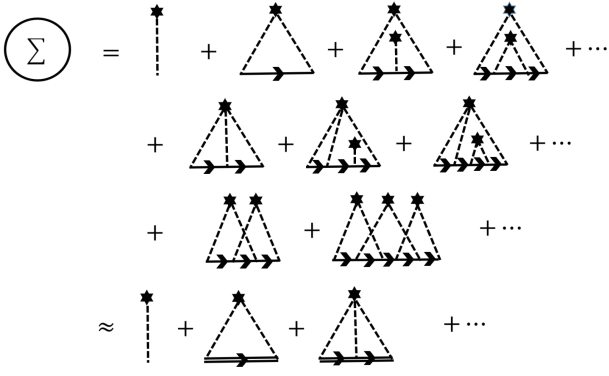}
\caption{
Self-energy diagrams for an impurity-scattered electron
} 
\label{fig:impurity self-energy}
\end{figure}

\begin{figure}[htb]
\includegraphics[width=1\linewidth,height=1.6in]{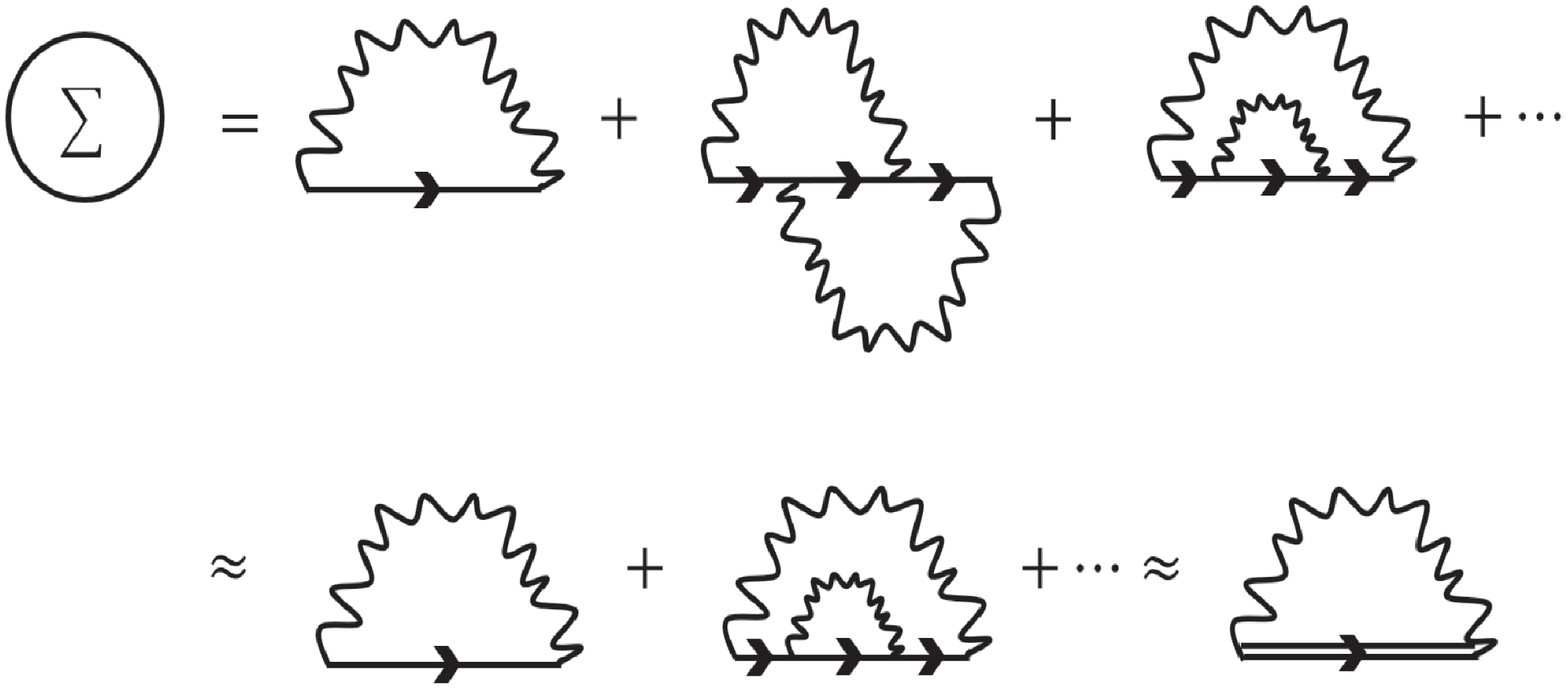}
\caption{
Self-energy diagrams for a phonon-scattered electron
} 
\label{fig:phonon self-energy}
\end{figure}

\section{Ward identities}
\label{app:Ward_identities}
\subsection{Elastic scattering}

The self-energy diagrams for an impurity-scattered electron are given by Fig.~\ref{fig:impurity self-energy}. The diagrams in the third line are negligible because the crossing induces the momentum restriction, which provides a reduction factor of the order \(\frac{1}{k_{\rm F}l}\), where \(l\) is the mean free path of the electrons. The first diagram in the second equality with the lowest-order perturbation only provides a constant energy shift, which can be absorbed in the chemical potential. As we only include the next lowest-order diagrams in the limit of $n_{\rm imp}\rightarrow 0$ into the ladder vertex correction, the proper self-energy reduces to the second diagram in the second equality given by

\begin{eqnarray}
\label{eq:impurity self energy}
\Sigma_{\alpha}(\bm{k},i\omega_n)\!
=n_{\rm imp}\!\sum_{\alpha'}\!\int\!\! \frac{d^dk'}{(2\pi)^d}|V_{\alpha',\bm{k}';\alpha,\bm{k}}|^2 \mathcal{G}_{\alpha'}(\bm{k}',i\omega_{n}).
\nonumber\\
\end{eqnarray}

The Ward identity can be obtained by subtracting \(\Sigma_{\alpha}(\bm{k},i\omega_n)\) from \(\Sigma_{\alpha}(\bm{k}+\bm{p},i\omega_n+ip_r)\) \cite{Mahan2000}:

\begin{eqnarray}
\label{eq:self energy subtraction}
&&\Sigma_{\alpha}(\bm{k}+\bm{p},i\omega_n+ip_r)-\Sigma_{\alpha}(\bm{k},i\omega_n) \nonumber\\
&&=n_{\rm imp}\sum_{\alpha'}\int\frac{d^dk'}{(2\pi)^d}|V_{\alpha',\bm{k}';\alpha,\bm{k}}|^2
\nonumber\\&&\times
\mathcal{G}_{\alpha'}(\bm{k}'+\bm{p},i\omega_{n}+ip_r)\mathcal{G}_{\alpha'}(\bm{k}',i\omega_{n}) \nonumber\\
&&\times [\xi_{\alpha',\bm{k}'+\bm{p}}-\xi_{\alpha',\bm{k}'}+\Sigma_{\alpha'}(\bm{k}'+\bm{p},i\omega_n+ip_r)
\nonumber\\&&
-\Sigma_{\alpha'}(\bm{k}',i\omega_n)-ip_r ].
\end{eqnarray}
Let us set \(ip_r\) to zero and take the limit \(\bm{p}\rightarrow 0\):

\begin{eqnarray}
\label{eq:self-energy k derivative1}
&&\left[\frac{\partial\Sigma_{\alpha}(\bm{k}'',i\omega_n)}{\partial \bm{k}''}\right]_{\bm{k}''=\bm{k}}\cdot \bm{p} \nonumber \\
&&=n_{\rm imp}\sum_{\alpha'}\int\frac{d^dk'}{(2\pi)^d}|V_{\alpha',\bm{k}';\alpha,\bm{k}}|^2 \mathcal{G}_{\alpha'}(\bm{k}',i\omega_{n})^{2} \nonumber\\
&&\times \left(\bm{v}_{\alpha',\bm{k}'}+\left[\frac{\partial\Sigma_{\alpha'}(\bm{k}'',i\omega_n)}{\partial \bm{k}''}\right]_{\bm{k}''=\bm{k}'}\right)\cdot \bm{p}.
\end{eqnarray}
Subsequently, let us take the limit \(i\nu_m\rightarrow 0\) in Eq.~(\ref{eq:vertex_dyson's equation}). As the integral equations in Eq.~(\ref{eq:self-energy k derivative1}) and the self-consistent Dyson's equation describing the vertex correction have the same form, we obtain the Ward identity given by
\begin{eqnarray}
\label{eq:ward identity}
   v_{\alpha,\bm{k}}^{(j)}\Lambda^{(j)}_{\alpha}(\bm{k},i\omega_n,i\omega_n)=v^{(j)}_{\alpha,\bm{k}}+\frac{\partial\Sigma_{\alpha}(\bm{k},i\omega_n)}{\partial k^{(j)}},
\end{eqnarray}
proving that the Ward identity still holds in anisotropic multiband systems for elastic scattering.

\subsection{Inelastic scattering}

As in the case of elastic impurity scattering, we show that the Ward identity holds for inelastic phonon scattering. The self-energy diagrams for a phonon-scattered electron are given by Fig.~\ref{fig:phonon self-energy}. The second diagram in the first equality can be regarded as the vertex correction to the first diagram, and such terms can be ignored according to Migdal's theorem \cite{Migdal1958}. Thus, the self-energy can be approximated as the diagram with self-consistency shown in the third equality given by
\begin{eqnarray}
\label{eq:phonon self energy}
  \Sigma_{\alpha}(\bm{k},i\omega_n)&&=-\frac{1}{\beta}\sum_{\alpha',\lambda}\sum_{iq_l}\int\frac{d^dq}{(2\pi)^d}|\matrixel{\alpha',\bm{k}+\bm{q}}{M_\lambda}{\alpha,\bm{k}}|^2
  \nonumber\\&&\times  D(\bm{q},iq_{l})\mathcal{G}_{\alpha'}(\bm{k}+\bm{q},i\omega_{n}+iq_{l}).
\end{eqnarray}

Repeating similar steps from Eq.~(\ref{eq:impurity self energy}) to Eq.~(\ref{eq:ward identity}) for impurity scattering, we have
\begin{eqnarray}
\label{eq:self-energy k derivative2}
&&\left[\frac{\partial\Sigma_{\alpha}(\bm{k}',i\omega_n)}{\partial \bm{k}'}\right]_{\bm{k}'=\bm{k}}\cdot \bm{p}
\nonumber\\&&=-\frac{1}{\beta}\sum_{\alpha',\lambda}\sum_{iq_l}\int\frac{d^dq}{(2\pi)^d}|\matrixel{\alpha',\bm{k}+\bm{q}}{M_\lambda}{\alpha,\bm{k}}|^2
\nonumber\\&&\times D(\bm{q},iq_{l})\mathcal{G}_{\alpha'}(\bm{k}+\bm{q},i\omega_{n}+iq_{l})^{2}
\nonumber\\&&\times \left(\bm{v}_{\alpha',\bm{k}+\bm{q}}+\left[\frac{\partial\Sigma_{\alpha'}(\bm{k}',i\omega_n)}{\partial \bm{k}'}\right]_{\bm{k}'=\bm{k}+\bm{q}}\right)\cdot \bm{p}, 
\end{eqnarray}
and
\begin{eqnarray}
\label{eq:vertex_zero freq limit2}
    &&v_{\alpha,\bm{k}}^{(j)}\Lambda^{(j)}_{\alpha}(\bm{k},i\omega_n,i\omega_n)
    \nonumber\\&&=v_{\alpha,\bm{k}}^{(j)}-\frac{1}{\beta }\sum_{\alpha',iq_l,\lambda}\int\frac{d^dq}{(2\pi)^d}|\matrixel{\alpha',\bm{k}+\bm{q}}{M_\lambda}{\alpha,\bm{k}}|^2
    \nonumber\\&&\times D(\bm{q},iq_l)v_{\alpha',\bm{k}+\bm{q}}^{(j)}  \Lambda^{(j)}_{\alpha'}(\bm{k}+\bm{q},i\omega_n +iq_l,i\omega_n +iq_l)
    \nonumber\\&&\times \mathcal{G}_{\alpha'}(\bm{k}+\bm{q},i\omega_n +iq_l)^2.
\end{eqnarray}
Thus, by comparing Eq.~(\ref{eq:self-energy k derivative2}) and Eq.~(\ref{eq:vertex_zero freq limit2}), we obtain 
\begin{eqnarray}
\label{eq:ward identity2}
   &&v_{\alpha,\bm{k}}^{(j)}\Lambda^{(j)}_{\alpha}(\bm{k},i\omega_n,i\omega_n)=v^{(j)}_{\alpha,\bm{k}}+\frac{\partial\Sigma_{\alpha}(\bm{k},i\omega_n)}{\partial k^{(j)}},
\end{eqnarray}
proving that the Ward identity still holds in anisotropic multiband systems for inelastic scattering.

\end{document}